\documentclass[letter,longauth,traditabstract]{aa} 
\usepackage{graphicx}
\usepackage{txfonts}
\usepackage{natbib} 
\usepackage{subfigure}

\bibpunct{(}{)}{;}{a}{}{,} 

\newcommand\micron{\mbox{$\mu$m}}%

\begin{document}
   \title{Cold Dust in Three Massive Evolved Stars in the LMC\thanks{Herschel is an ESA space observatory with
science instruments provided by European-led Principal Investigator
consortia and with important participation from NASA.}}

   \author{M. L. Boyer\inst{1}     
     \and B. Sargent\inst{1}
     \and J.~Th. van Loon\inst{2}
     \and S. Srinivasan\inst{3}
     \and G. C. Clayton\inst{4}
     \and F. Kemper\inst{5}
     \and L.~J. Smith\inst{1}
     \and M. Matsuura\inst{6,7}
     \and Paul M. Woods\inst{5}
     \and M. Marengo\inst{8}
     \and M. Meixner\inst{1}\fnmsep\thanks{Visiting Scientist at Smithsonian Astrophysical Observatory, Harvard-CfA, 60 Garden St., Cambridge, MA, 02138}
     \and C. Engelbracht\inst{9}
     \and K.~D. Gordon\inst{1}
     \and S. Hony\inst{10}
     \and R. Indebetouw\inst{11}
     \and K. Misselt \inst{9}
     \and K. Okumura\inst{10}
     \and P. Panuzzo\inst{10}
     \and D. Riebel\inst{12}
     \and J. Roman-Duval\inst{1}
     \and M. Sauvage\inst{10}
     \and G.~C. Sloan\inst{13} 
  }

 \institute{Space Telescope Science Institute, 3700 San Martin Drive, Baltimore, MD 21218, USA  \email{mboyer@stsci.edu}
\and School of Physical \& Geographical Sciences, Lennard-Jones Laboratories, Keele University, Staffordshire ST5 5BG, UK
\and Institut d'Astrophysique de Paris, CNRS UPR 341, 98bis, Boulevard Arago,  Paris, F-75014  
\and Louisiana State University, Department of Physics \& Astronomy, 233-A Nicholson Hall, Tower Dr., Baton Rouge, LA 70803, USA
\and Jodrell Bank Centre for Astrophysics, Alan Turing Building, School of Physics and Astronomy, University of Manchester, Oxford Road, Manchester  M13 9PL, UK  
\and  Department of Physics and Astronomy, University College London, Gower Street, London WC1E 6BT, UK 
\and Mullard Space Science Laboratory, University College London, Holmbury St. Mary, Dorking, Surrey RH5 6NT, UK
\and  Department of Physics and Astronomy, Iowa State University, Ames, IA, 50011 USA
\and Steward Observatory, University of Arizona, 933 North Cherry Ave., Tucson, AZ 85721, USA 
\and CEA, Laboratoire AIM, Irfu/SAp, Orme des Merisiers, F-91191 Gif-sur-Yvette, France
\and National Radio Astronomy Observatory, Department of Astronomy, University of Virginia, PO Box 3818, Charlottesville, VA 22903 USA
\and Johns Hopkins University, Department of Physics and Astronomy, Homewood Campus, Baltimore, MD 21218, USA  
\and Department of Astronomy, Cornell University, Ithaca, NY 14853, USA
}

   \date{Received $<$date$>$ / Accepted $<$date$>$}

  \abstract {Massive evolved stars can produce large amounts of dust,
  and far-infrared (IR) data are essential for determining the
  contribution of cold dust to the total dust mass. Using $Herschel$,
  we search for cold dust in three very dusty massive evolved stars in
  the Large Magellanic Cloud: R71 is a Luminous Blue Variable,
  HD\,36402 is a Wolf-Rayet triple system, and IRAS05280--6910 is a
  red supergiant. We model the spectral energy distributions using
  radiative transfer codes and find that these three stars have mass-loss rates up to $10^{-3}~M_\odot~{\rm yr}^{-1}$, suggesting that
  high-mass stars are important contributors to the life-cycle of
  dust. We found far-IR excesses in two objects, but these excesses
  appear to be associated with ISM and star-forming regions. Cold dust
  ($T<100$~K) may thus not be an important contributor to the dust
  masses of evolved stars.}

   \maketitle
   \titlerunning{Cold dust in Extragalactic Supernova Progenitors}
   \authorrunning{M.L.\,Boyer et al.}
%

\section{Introduction}

Intermediate-mass asymptotic giant branch (AGB) stars are potentially
the dominant dust source in the Galaxy \citep{gehrz89} and in
low-metallicity environments like the Large Magellanic Cloud
\citep[LMC;][]{srinivasan09, matsuura09} and other dwarf galaxies
\citep{boyer09_dirr}. However, dust production in high-mass stars
($\gtrsim$$8~M_\odot$) remains uncertain. It has been suggested that
supernovae (SNe) might be the dominant dust factory at high-$z$, since
intermediate-mass stars have not yet had time to evolve into AGB stars
\citep{morgan03, dwek09}. However, the amount of dust formed in SNe in
the local universe is much less than required to explain the dust seen
at high-$z$ \citep[e.g.,][]{sugerman06,andrews10}. It is also unclear
if the dust forms before or after the SN explosion.  Alternatively,
\citet{sloan09} and \citet{valiante09} show that AGB stars {\it can}
contribute dust at high redshifts.  These studies point to a need to
measure the total dust mass from all types of stars to obtain a global
picture of dust evolution in galaxies.

Part of the LMC was observed with {\it Herschel} \citep[][this volume]{pilbratt} as part of the science demonstration program
(SDP) and the Legacy program entitled {\it A HERschel Inventory of The
Agents of Galaxy Evolution} \citep[HERITAGE;][this volume]{meixner}. In this letter, we describe a first look at {\it
Herschel} Photodetector Array Camera and Spectrometer \citep[PACS;][this volume]{poglitsch} and Spectral and Photometric Imaging
REceiver \citep[SPIRE;][this volume]{griffin} detections of 3
examples of dust-producing massive evolved stars in the LMC: the
Wolf-Rayet (WR) system HD\,36402, the luminous blue variable (LBV)
HDE\,269006 (or R71), and the red supergiant (RSG) OH/IR star
IRAS05280--6910. While such stars have been studied extensively in the
mid-IR \citep[e.g.,][]{morrisiso99, clark03, crowther07,
bonanos09, vanloon10}, this study is among the first to probe them at
$\lambda\gtrsim200~\micron{}$.

\begin{table*}%
\caption{Target Information and {\it Herschel} Flux Densities}
\label{tab:targets}
\centering
\begin{tabular}{llllllll}
\hline
\hline
Name & R.A., Dec. (J2000) & Type & \multicolumn{5}{c}{$F_\nu$ (mJy)}\\ 
& & & 100~\micron & 160~\micron & 250~\micron & 350~\micron & 500~\micron \\
\hline
R71 & 05$^{\rm h}$02$^{\rm m}$07.39$^{\rm s}$, $-$71${\degr}$20${\arcmin}$13.1${\arcsec}$  & LBV & $447\pm8$ & $105\pm10$ & $50\pm6$ & $49\pm13$ & $27\pm9$\\
IRAS05280$-$6910 & 05$^{\rm h}$27$^{\rm m}$40.11$^{\rm s}$, $-$69${\degr}$08${\arcmin}$04.5${\arcsec}$ & RSG, OH/IR &...&...& $205\pm8$ & $46\pm13$ & $<$$11\pm10$ \\
HD\,36402 & 05$^{\rm h}$26$^{\rm m}$03.94$^{\rm s}$, $-$67${\degr}$29${\arcmin}$57.9${\arcsec}$ & WC4($+$O?)$+$O8I &...&...&...&...&...\\
HD\,36402 IR1/YSO2& 05$^{\rm h}$26$^{\rm m}$02.99$^{\rm s}$, $-$67${\degr}$29${\arcmin}$57.7${\arcsec}$ &YSO?& $531\pm10$ & $872\pm33$ & $677\pm33$ & $<$$431\pm23$ & $<$$132\pm21$\\
\hline
\end{tabular}
\begin{list}{}{}
\item[{\sc Notes.}--] Quoted uncertainties are 1\,$\sigma$. Upper
  limits are either below the 3\,$\sigma$ detection level or include
  substantial flux from an adjacent unresolved source. IRAS05280--6910
  does not fall within the PACS coverage.  The WR star itself is not
  resolved in {\it Herschel} images; it is severely blended with an IR
  source (IR1/YSO2).  The 350 and 500~\micron{} fluxes of IR1/YSO2
  include a nearby YSO (N51-YSO1), which is unresolved from IR1/YSO2
  at these wavelengths. Aperture corrections for PACS and SPIRE were
  estimated using the current point-spread functions, and are roughly
  1.4 and 1.2, respectively. None of the sources are heavily affect by
  high and/or variable backgrounds (Fig.~\ref{fig:raw}).
\end{list}
\end{table*}

\begin{figure}
\centering
\hbox{
\includegraphics[width=0.155\textwidth]{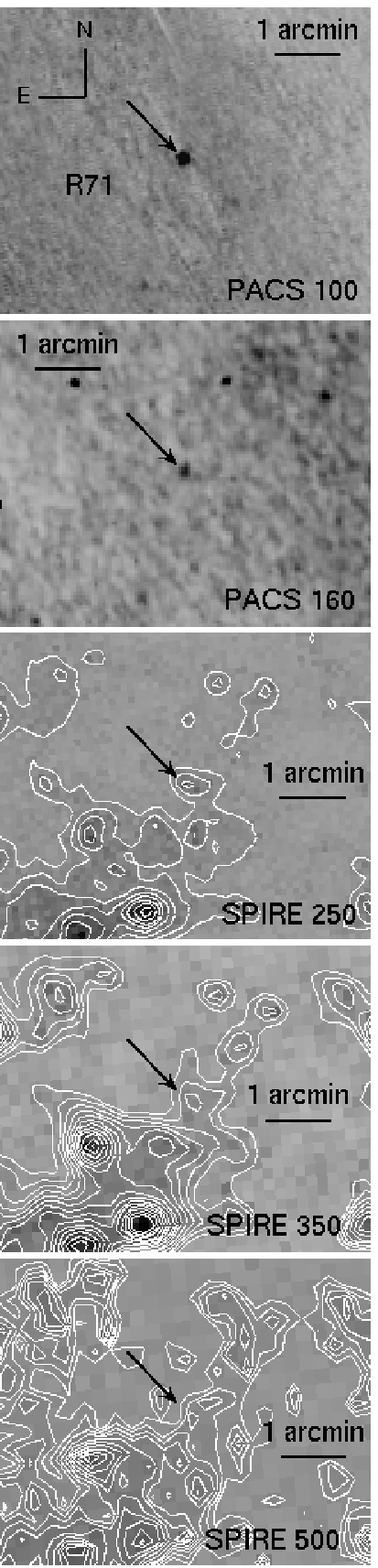}
\includegraphics[width=0.155\textwidth]{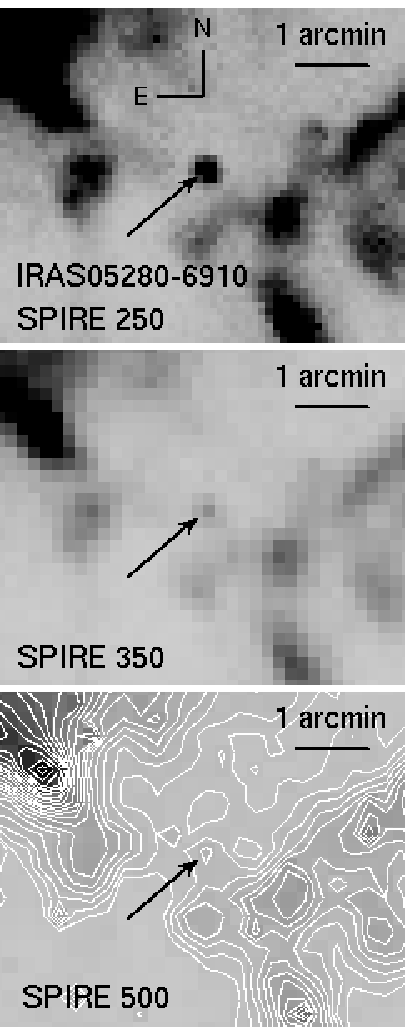}
\includegraphics[width=0.155\textwidth]{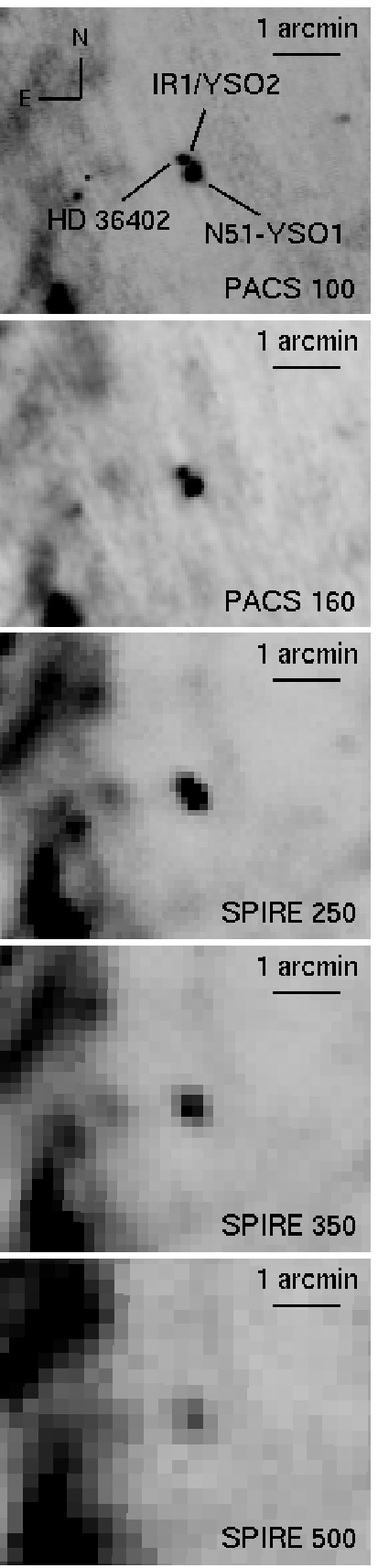}
}
\caption{{\it Herschel} images of R71 (left panels), IRAS05280-6910
  (middle panels), and HD\,36402 (right panels, also see
  Fig.~\ref{fig:img}). Contours on a linear scale are
  included where it is difficult to see the detection. \label{fig:raw}}
\end{figure}

\section{Mid- to far-infrared photometry}
\label{sec:data}
We obtained PACS (100 and 160~\micron{}) and SPIRE (250, 350 and
500~\micron{}) fluxes using apertures roughly the size of the
source full-width at half-maximum and sky apertures
avoiding regions of high background (Table~\ref{tab:targets}).  We also performed aperture
photometry on the Multiband Imaging Photometer for {\it Spitzer}
\citep[MIPS;][]{rieke04} images from the {\it Surveying The Agents of
Galaxy Evolution Spitzer} Legacy program
\citep[SAGE;][]{meixner06}.

In Sect.\,\ref{sec:seds} we examine the optical to far-IR spectral energy
distributions (SEDs). Optical $UBVI$ and near-IR $JHK$ photometry are
from the Magellanic Clouds Photometric Survey \citep{zaritsky97} and
the Two Micron All-Sky Survey \citep{skrutskie06}, via the SAGE
catalog.  $J$, $K$, and $L'$ photometry of IRAS05280--6910 are from
\citet{vanloon05b}. $I-$band photometry for HD\,36402 is from the Deep
Near-Infrared Southern Sky Survey \citep{epchtein97}. InfraRed Array
Camera \citep[IRAC;][]{fazio04} $3.6 - 8.0$~\micron{} photometry is
from the SAGE catalog. Spectra from the InfraRed Spectrograph
\citep[IRS; $5.2 - 38$~\micron{};][]{houck04} and MIPS spectra
(MIPS-SED; $52 - 97$~\micron{}) are also included \citep[e.g.,
SAGE-Spec;][]{ kemper10,vanloon10}. We correct for extinction using
the extinction map from \citet{schlegel98}, assuming the stars lie
midway the LMC contribution: $A_{\rm V} \approx 1$~mag for HD\,36402
and IRS05280-6910 and $A_{\rm V} \approx 0.4$~mag for R71.

\section{The IR nature of detected sources}
\label{sec:seds}

The vast majority of point-sources detected in the LMC SDP data appear
to be young stellar objects (YSOs) or background galaxies
\citep[][this volume]{sewilo}; AGB stars were not detected. Here, we
describe the {\it Herschel} observations of three very dusty massive
stars (Fig.~\ref{fig:raw}).

\begin{figure}
\centering

\includegraphics[width=0.46\textwidth]{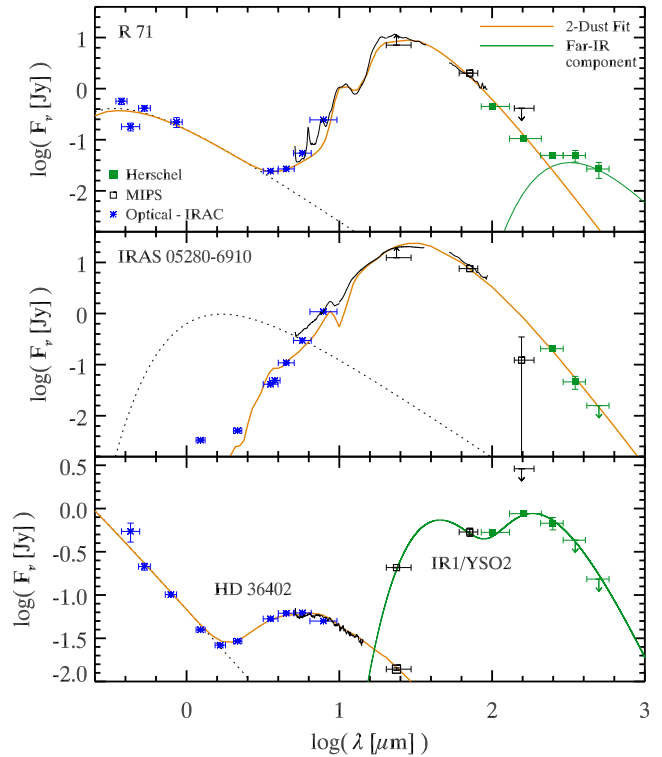}

\caption{SEDs of R71 (top), IRAS05280--6910 (middle) and HD\,36402
  (bottom), fit to 2-D{\sc ust} models (orange lines).  Far-IR excesses are
  fit to modified blackbodies (green lines). The IRS and MIPS-SED
  spectra are shown in black and the stellar components are the dotted
  lines. In the upper two panels, the 24~\micron{} points are
  saturated (lower limits). In the bottom panel, the bright far-IR
  emission originates from IR1/YSO2; the WR system is detected in
  PACS, but unresolved from IR1. See text.
\label{fig:seds}}
\end{figure}

\subsection{R71}
\label{sec:r71}

R71 (Fig.~\ref{fig:raw}) is an LBV with $M_{\rm birth} \approx
40~M_\odot$ \citep{lennon93}, and is currently experiencing an
unprecedented eruption that began in 2005. One month prior to the SDP
observations (Oct 2009), it showed a $\sim$2~mag increase in visual
brightness and the $V$-band light curve was just beginning to plateau
\citep{szczygiel10}. Van Loon et al. (2010) find a lack of cold dust
in R71, as indicated by {\it Spitzer} data ($T_{\rm d} \lesssim
100~{\rm K}$).  The {\it Spitzer} data were acquired just prior to the
current outburst, and the {\it Herschel} data presented here were
obtained at near-maximum. However, the PACS and MIPS points appear
consistent with each other, such that a model fit to the optical to
MIPS data also agrees with the PACS points. This indicates that the
increased emission from the photosphere has not yet significantly
affected the 100 -- 160~\micron{} flux. Note that the MIPS
160~\micron{} point is an upper limit.

IRS and {\it ISO} spectra of R71 show strong PAH, crystalline forsterite and enstatite features
\citep{voors99,morris08,buchanan09,waters10}. A 10.5~\micron{} amorphous silicate feature indicates dominantly
oxygen-rich (O-rich) chemistry. \citet{morris08} and \citet{voors99}
speculate that the dust was formed during a prior RSG phase.

Figure~\ref{fig:seds} shows R71's SED. Ultraviolet spectra
\citep{blair09} are consistent with the stellar component included in
our model. We fit the SED with a 2-D{\sc ust} model
\citep[online Table~\ref{tab:fit};][]{ueta03}, adopting spherical symmetry
and grain properties of the dominant species: amorphous
silicate. Three dust components are visible, including a previously
unknown excess visible in the SPIRE data at 250, 350 and 500~\micron{}
(8, 3.8, and 3\,$\sigma$ detections, respectively).  The two warmer
components are modeled by the 2-D{\sc ust} code using two concentric
dust shells. We assume the gas-to-dust ratio for the LMC is $\psi=300$
\citep[cf.][this volume]{meixner}, yielding $\dot{M} \sim
10^{-6}~M_\odot~{\rm yr}^{-1}$ for the inner dust shell and $\dot{M}
\sim 10^{-3}~M_\odot~{\rm yr}^{-1}$ for the outer shell, assuming
$\upsilon_{\rm wind} =$ 10~km\,s$^{-1}$, consistent with the wind
speed of a shell ejected during an RSG phase.  \citet{stahl86} find
$\upsilon \approx 160$~km\,s$^{-1}$ from the H$\alpha$ line profile,
indicating the MLR for the inner shell could be an order of magnitude
higher. A {\sc dusty} model \citep{nenkova99} fits the SED equally
well and estimates a MLR that is the 2-D{\sc ust} outer
shell value.

We fit the far-IR excess with a modified blackbody:
$F_\lambda~\propto~B_\lambda\left(T_{\rm d}\right) \left(1 -
e^{-\tau_\lambda}\right)$, where $B_\lambda(T_{\rm d})$ is the Planck
function at temperature $T_{\rm d}$, $\tau_\lambda$ is the optical
depth, and $\tau_\lambda \propto \lambda^{-\beta}$. Here, we use
$\beta=1.5$. The resulting dust temperature is $9 \pm 1$~K, which is
extremely low compared to the expected temperature of outer shells in
evolved stars \citep[$\sim$30\,K;][]{speck00}, and is instead
consistent with temperatures of dense ISM dust clouds (see
Sect.\,\ref{sec:disc}).

\onltab{2}{

\begin{table*}
\caption{SED Fit Parameters and Results}
\label{tab:fit}
\centering
\begin{tabular}{llllllllll}
\hline
\hline

&$T_*$&$L_*$&$T_{\rm in}$&$R_*$&$R_{\rm out}/R_{\rm in}$&$R_{\rm in}$&
$\dot{M}_{\rm total}$& $T_{\rm dust}^{\rm sub-mm}$ & $M_{\rm dust}^{\rm sub-mm}$ \\

&($10^4$\,K)&($10^5\,L_\odot$)&(K)& ($R_\odot$) & & (km) & ($M_\odot~{\rm yr}^{-1}$)& (K) & ($M_\odot$) \\

\hline
R71 (inner/outer shell)& $1.5$ & $4.6$ &490/120&100&2/1.6&$\sim$$10^{11}$/$10^{12}$&$\sim$$10^{-6}$ / $\sim$$10^{-4}$ & $9 \pm 1$ & $\gtrsim$$10^{-1}$ \\

IRAS05280$-$6910&0.3&$2.2$&250&1700&30&$\sim$$10^{11}$&$\sim$$10^{-3}$ &...&... \\ 

HD\,36402& $18$ & $4.6$ &960&15&300&$\sim$$10^{10}$&$7\times10^{-6}$ &...&... \\

HD\,36402 IR1 (YSO2)&...&...&...&...&...&...&...&$64$/$15$ ($\pm 1$) & $10^{-12}$/$>$$1.5 \pm 0.4$\\ 
\hline
\end{tabular}
\begin{list}{}{}
\item[{\sc Notes}.--] Fitting results from 2-D{\sc ust}.  {\sc dusty}
fits for R71 and IRAS05280-6910 estimate outer shell MLRs 2$\times$ to
4$\times$ less than the 2-D{\sc ust} results.  IR1/YSO2 was fit with
two modified blackbodies (64~K and 15~K). For R71, we use O-deficient
silicate optical constants \citep{ossenkopf92} at long wavelengths and
astronomical silicate optical constants \citep{draine84} at $\lambda <
0.18$~\micron{}. For IRAS05280--6910, we use \citet{ossenkopf92}
O-deficient silicates and for HD\,36402, we use amorphous carbon
grains from \citet{zubko96}. The 2-D{\sc ust} models use a KMH grain
size distribution \citep*{kim94} with $a_{\rm min}=0.01\,\micron$ and
$a_0=0.1\,\micron$ ($a_0=1~\micron$ for IRAS05280-6910). Stellar
parameters for R71 are consistent with those from
\citet{lennon93}. $T_{\rm dust}^{\rm sub-mm}$ is the temperature of
the coldest dust component. For IRAS5280-6910, $\upsilon_{\rm
wind}$ is measured from the maser emission
\citep[20~km\,s$^{-1}$;][]{marshall04}. For R71, we use
10~km\,s$^{-1}$, but the velocity measured from the H$\alpha$ profile
indicates the inner shell velocity may be $>$10$\times$ larger
\citep{stahl86}. For HD36402, a typical velocity for a WC4 star is
assumed \citep[3000~km\,s$^{-1}$;][]{willis04}.

\end{list}
\end{table*}
}

\begin{figure}
\centering

\subfigure[]{\includegraphics[width=0.24\textwidth]{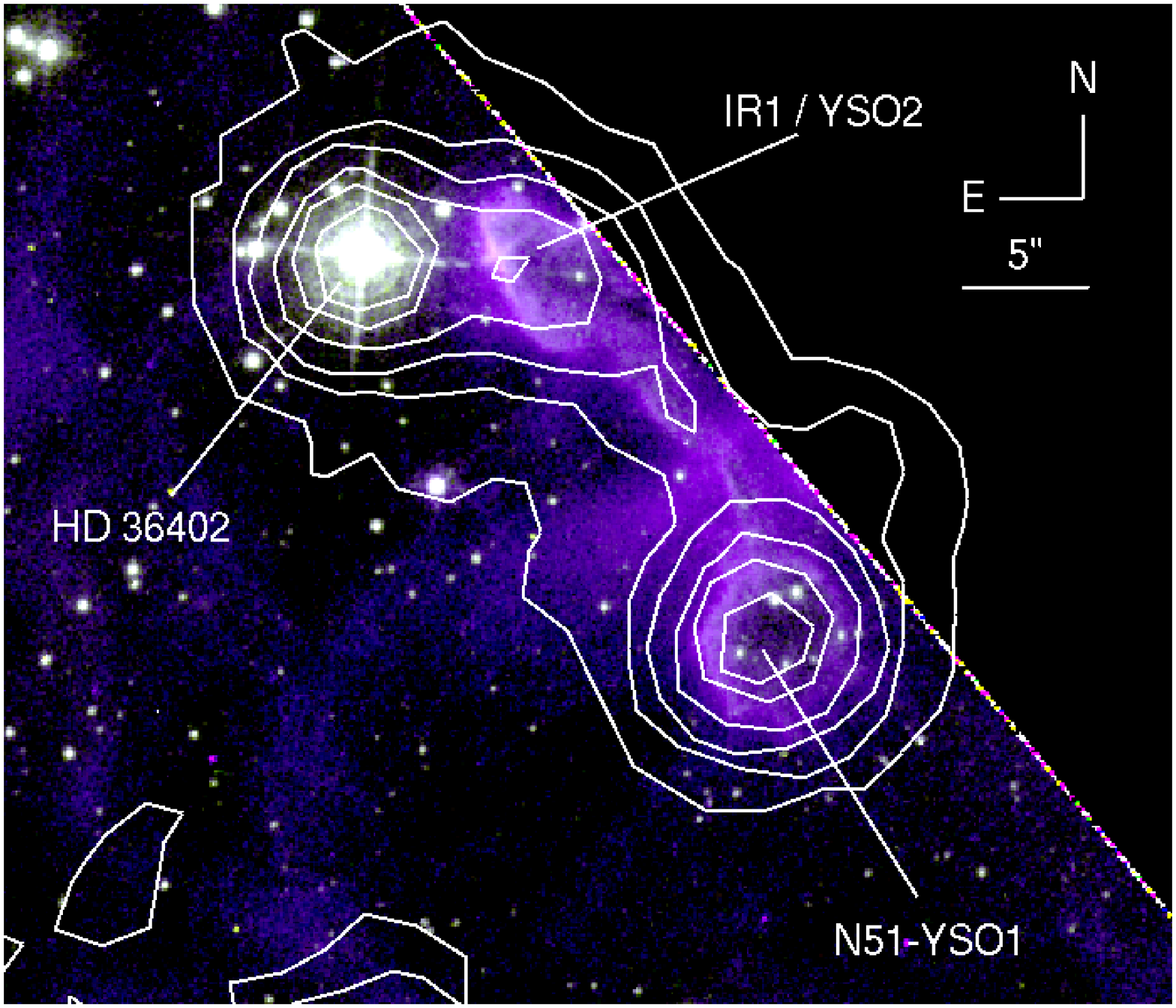}}
\subfigure[]{\includegraphics[width=0.24\textwidth]{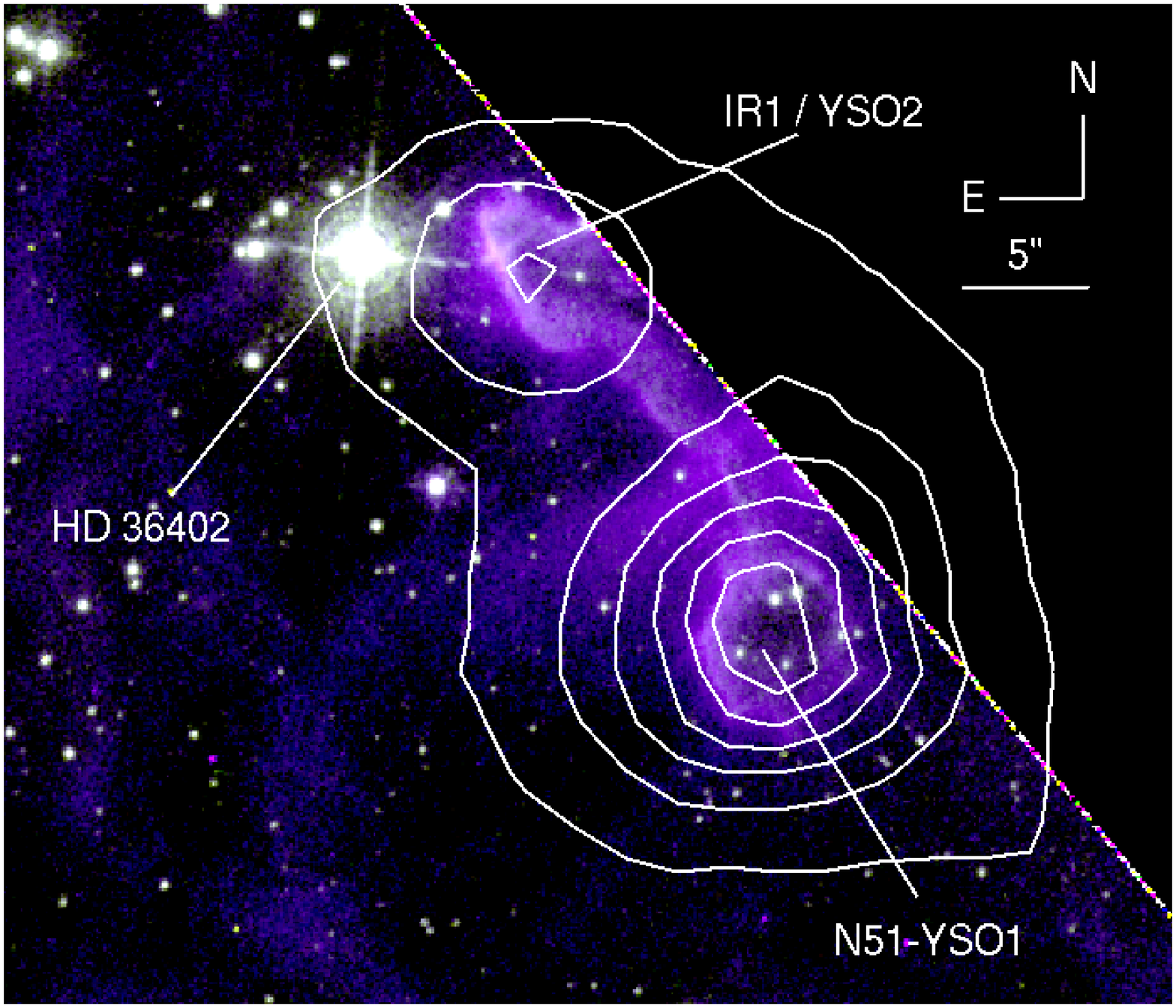}}
\caption{Composite {\it HST}/WFPC2 image of HD\,36402 in H$\alpha$ and
[\ion{S}{II}] (purple). The contours represent (a) IRAC 8~\micron{}
and (b) PACS 100~\micron{}. \label{fig:img}}
\end{figure}

\subsection{IRAS05280--6910}
\label{sec:iras}

IRAS05280--6910 is an RSG OH/IR star. It is detected in all {\it
Spitzer} and {\it Herschel} bands up to 350~\micron{}
($>$3.5\,$\sigma$; Figs.~\ref{fig:raw} and \ref{fig:seds}). The star is heavily
extinguished in the optical and near-IR, and the 10 and 18~\micron{}
silicate features are seen in absorption in the IRS spectrum
\citep{kemper10}. Van Loon et al. (2010) find no indication of dust
colder than $\sim$$100$~K in the MIPS data.

We fit the SED of IRAS05280--6910 using 2-D{\sc ust} (flattened
  geometry) and {\sc dusty} (spherical symmetry).  Both models fit the
  SED reasonably well (only the 2-D{\sc ust} model is shown in
  Fig.~\ref{fig:seds}), but predict stronger absorption in the
  10~\micron{} silicate feature than is seen in the IRS spectrum, even
  when including grains as large as 1~\micron{} (increasing $a_0$
  suppresses the silicate feature). This discrepancy may be due to the
  inclusion in the slit of a nearby RSG (WOH\,G347), which is too
  faint to contribute to the SED at other wavelengths, but shows
  emission near 10~\micron{} with enough flux to veil the silicate
  absorption by the necessary amount \citep{vanloon05b}.  Based on
  both model fits and assuming an outflow velocity of
  $\approx$20~km\,s$^{-1}$ \citep[as measured from maser
  emission;][]{marshall04}, the mass-loss rate (MLR) was $\dot{M} =
  2-8 \times 10^{-3}~M_\odot~{\rm yr}^{-1}$ when the dust was produced
  (online Table~\ref{tab:fit}). There is no evidence of excess emission at
  $\lambda > 100$~\micron{} in the SDP data, implying there is no
  significant contribution from cold ($\ll 100$~K) and/or large
  (\micron{} -- mm size) grains.

\subsection{HD\,36402}
\label{sec:bat}

HD\,36402 (Fig.~\ref{fig:img}) is a WR star that is part of a triple
system \citep{moffat90}. It is the reddest LMC WR star studied by
\citet{bonanos09} in IRAC, potentially due to dust formed in colliding
stellar winds \citep{crowther07}.  HD\,36402 is detected by PACS
(Fig. 3b), which might indicate the presence of some cool
dust. However, the system is almost totally unresolved from the stronger
far-IR emission immediately to the west (Figs.~\ref{fig:raw} and
\ref{fig:img}b), which appears to originate from a nearby molecular
cloud, visible in H$\alpha$ \citep[Fig.~\ref{fig:img};][]{dopita94,
chu05}. We refer to this emission as HD\,36402 IR1.

The IRS spectrum of HD\,36402 from SAGE-Spec is featureless and
reminiscent of R Coronae Borealis stars, which are C-rich (no silicate
dust) and hydrogen-deficient (no PAHs). The continuum emission is
thought to be due to amorphous carbon dust
\citep[e.g.,][]{kraemer05}. The SED (Fig.~\ref{fig:seds}) shows a stellar component and a component from a detached, dusty shell
($\sim$$2-24$~\micron). The far-IR SED is dominated by IR1,
so an accurate fit to the WR system in the far-IR is not
possible with this dataset. In the lowest resolution images (MIPS 160,
SPIRE 350 and 500~\micron{}), IR1 is also unresolved from N51-YSO1
(upper limits in Fig.~\ref{fig:seds}), further
complicating the SED. A 2-D{\sc ust} model (Fig.~\ref{fig:seds},
orange line) gives $\dot{M} = 7 \times 10^{-6}~M_\odot~{\rm yr}^{-1}$.

What is the nature of IR1?  Based on its IRAC colors, \citet{chu05}
identify it as a YSO (N51-YSO2). We have attempted to fit a
two-component modified blackbody ($\beta = 2.0$) to the IR1 SED to
check if it instead originates from dust heated directly by the
radiation emanating from HD\,36402. These fits yield a dust
temperature of 64~K, with 15~K dust in its wake, consistent with this
scenario.  However, a temperature of 64~K places the molecular
cloud $\sim$$0.3 \pm 0.1$~pc from the WR system, assuming typical
luminosities of a WC4 star and two O star companions.  The
high-resolution H$\alpha$ image (Fig.~\ref{fig:img}) puts the
molecular cloud at least 1.2~pc away from the WR system, assuming a
distance of 50~kpc to the LMC \citep{schaefer08}.  It thus seems
likely that the far-IR emission of IR1 is indeed related to a YSO
embedded in the molecular cloud rather than to direct heating from
HD\,36402. In the {\it HST} image (Fig.~\ref{fig:img}), the cloud
appears brightest in the region adjacent to HD\,36402, which may
indicate {\it some} interaction between the two. If this is indeed the
case, it is plausible that the formation of YSO2 was triggered by
HD\,36402.

\section{Implications}
\label{sec:disc}

The new far-IR {\it Herschel} data have allowed us to take the first
steps in assessing the contribution of cold dust to the total dust
mass in 3 massive stars. For the RSG IRAS05280--6910, the models give
a total dust mass of $\approx$$0.3~M_\odot$. This dust mass is
extremely large for an RSG star; indeed it is also highly uncertain,
as it is computed from the dust density distribution and the grain
composition and size distribution, which introduces many
degeneracies. By adjusting these parameters, we can reduce the dust
mass by at least 60\%, possibly more. The MLR (online Table~\ref{tab:fit}),
while still uncertain, is a more reliable quantity and suggests the
star has entered a necessarily-brief phase of extreme mass loss. The
large mass found using our simple, preliminary models may indicate a
complicated geometry with a preferential viewing angle requiring more
extensive modeling, or perhaps the RSG is embedded in a dusty
cloud. Whatever the total envelope mass, a lack of excess far-IR
emission over the model indicates that ancient, cold dust does not
contribute significantly.  Dust at 50~K, for instance, must
contribute less than 10\% of the dust mass implied by the model fit to
explain this lack of far-IR excess.

The LBV R71 shows three dust components.  A component emitting at
$\sim$$10 - 100~\micron$ resembles RSG dust. A second component
dominates at $3 - 10~\micron{}$. Its temperature ($T_{\rm in}=490$~K)
suggests it formed $<$$50$~yrs ago, assuming $\upsilon\sim100~{\rm
km~s}^{-1}$ and $T_{0} = 1000$~K, perhaps during the 1970s outburst
\citep{wolf86}.  The third, more tentative dust component is visible
at $\lambda>250$~\micron. If this feature corresponds to
cold circumstellar dust, the implied dust mass is $M_{\rm dust}
\gtrsim 10^{-1}~M_\odot$, following \citet{evans03} and assuming the
absorption coefficient, $\kappa(500\,\mu {\rm m})$, is $20~{\rm
cm^2~g}^{-1}$ \citep{ossenkopf94}. This is far too much dust for a
star of this mass; together with its very cold temperature (9~K), this
high mass suggests that the far-IR emission is pre-existing ISM dust
swept up by stellar winds and/or is ISM dust along the
line-of-sight. Indeed, the contours in Figure~\ref{fig:raw}
appear to show diffuse emission at the position of R71 at the longest
wavelengths. If the far-IR emission instead originates from the
circumstellar envelope, then very large grains similar to the cm-sized
grains in the Egg Nebula \citep{jura00} might explain the far-IR
emission without requiring the implied large dust mass and cold
temperature.  Follow-up spectroscopy or deeper SPIRE imaging may help
uncover the nature of the far-IR emission.

The WR system, HD\,36402, seems to be forming dust in colliding winds,
which is too warm to emit much at far-IR wavelengths. The apparent
far-IR emission from HD\,36402 (Fig. 3b) is unfortunately
totally overwhelmed by far-IR emission originating from a nearby
molecular cloud. Assuming $\kappa(70\,\micron) \approx 140~{\rm
cm^2~g}^{-1}$ and $\kappa(500\,\micron) \approx 20~{\rm cm^2~g}^{-1}$
\citep{ossenkopf94}, we find $M_{\rm dust}^{63\,\rm K} = 1.0 (\pm 0.3)
\times 10^{-2}~M_\odot$ and $M_{\rm dust}^{15\,\rm K} = 1.5 \pm
0.4~M_\odot$ of dust in the molecular cloud.

Stellar evolution models, while still uncertain, show that massive
stars like these eventually explode as SNe. Due to its proximity to a
molecular cloud, the HD\,36402 remnant may resemble the SN remnant
N49, which has swept up 0.2~$M_\odot$ of dust from the ISM
\citep[][this volume]{vanloon10,otsuka}. The absence of strong
evidence for very large grains in R71 and IRAS05280-6910 does not raise the
prospects of RSG dust surviving a SN blast. Regardless of their fates,
the dust masses in these 3 stars are quite large, compared with the
dust mass found in a typical AGB star, showing that high-mass stars
are important contributors to the life-cycle of dust even in
low-metallicity environments like the LMC. However, we emphasize
that we do not find strong evidence for cold dust and/or large grains
in any of the three objects discussed here, except where it is certain
(HD\,36402) or likely (R71) to be of interstellar origin and not
synthesized by the object itself.  These observations indicate that
far-IR data of a much larger sample of luminous evolved stars in both
Magellanic Clouds will be obtained in the full HERITAGE dataset, from
which we expect clearer patterns to emerge.

\begin{acknowledgements}

We thank the referee for his or her helpful comments. This publication
includes observations made with the NASA/ESA {\it HST}, and obtained
from the {\it Hubble} Legacy Archive, which is a collaboration between
the Space Telescope Science Institute (STScI/NASA), the Space
Telescope European Coordinating Facility (ST-ECF/ESA) and the Canadian
Astronomy Data Centre (CADC/NRC/CSA). We acknowledge financial support
from the NASA {\it Herschel} Science Center, JPL contract \# 1381522.
We thank the contributions and support from the European Space Agency
(ESA), the PACS and SPIRE teams, the {\it Herschel} Science Center and
the NASA {\it Herschel} Science Center (esp. A. Barbar and K. Xu) and
the PACS/SPIRE instrument control center at CEA-Saclay, which made
this work possible.
    \end{acknowledgements}

\bibliographystyle{aa} 
\bibliography{14513refs} 

\end{document}